\documentclass[journal]{IEEEtran}
\usepackage{amsmath,amsfonts}

\usepackage{cite}
\usepackage{array}

\usepackage{tabularx}
\usepackage{booktabs}

\usepackage{bbding}
\usepackage[caption=false,font=normalsize,labelfont=sf,textfont=sf]{subfig}
\usepackage{multirow}
\usepackage{makecell}
\usepackage{textcomp}
\usepackage{stfloats}
\usepackage{url}
\usepackage{verbatim}
\usepackage{graphicx}
\usepackage[T1]{fontenc} %摘要字体加粗
\usepackage{amssymb} % 三角等号
\usepackage{algorithmic}
\usepackage{algorithm}
\usepackage[backref=False]{hyperref}
\def\BibTeX{{\rm B\kern-.05em{\sc i\kern-.025em b}\kern-.08em
		T\kern-.1667em\lower.7ex\hbox{E}\kern-.125emX}}
\usepackage{balance}
\newcommand{\Rmnum}[1]{\uppercase\expandafter{\romannumeral #1}}  %定义命令输入大写罗马数字
\usepackage{tikz,xcolor,hyperref}

\usepackage{amsmath}
\usepackage{subcaption}
\usepackage{graphicx}
\usepackage{amssymb}

\definecolor{lime}{HTML}{A6CE39}
\DeclareRobustCommand{\orcidicon}{
	\begin{tikzpicture}
		\draw[lime, fill=lime] (0,0)
		circle[radius=0.16]
		node[white]{{\fontfamily{qag}\selectfont \tiny \.{I}D}}; 
	\end{tikzpicture}
	\hspace{-2mm}
}
\foreach \x in {A, ..., Z}{%
	\expandafter\xdef\csname orcid\x\endcsname{\noexpand\href{https://orcid.org/\csname orcidauthor\x\endcsname}{\noexpand\orcidicon}}
}

\begin{document}

\title{Defending Consensus-Based Cooperative Sensing Against Conformity-Driven Byzantine Attacks in 6G UAV Swarm Networks}

\author{Ruixing~Ren\hspace{-1.5mm}\orcidA{}, Junhui~Zhao\hspace{-1.5mm}\orcidB{},~\IEEEmembership{Senior~Member,~IEEE}, He~Fang\hspace{-1.5mm}\orcidC{},~\IEEEmembership{Member,~IEEE}, Xiaoke~Sun\hspace{-1.5mm}\orcidD{},\\ Jiamin~Li\hspace{-1.5mm}\orcidE{},~\IEEEmembership{Member,~IEEE}, Pengcheng~Zhu\hspace{-1.5mm}\orcidF{},~\IEEEmembership{Member,~IEEE}, and Dongming~Wang\hspace{-1.5mm}\orcidG{},~\IEEEmembership{Member,~IEEE},
\thanks{This work was supported in part by National Natural Science Foundation of China under Grant U25B2007, in part by the the Fundamental Research Funds for the central Universities under Grant 2025JBZX060, in part by National Engineering Research Center of System Technology for High-Speed	Railway and Urban Rail Transit under Grant 2024YJ255. (Corresponding author: Junhui Zhao.)}%
\thanks{Ruixing Ren, Junhui Zhao are with the School of Electronic and Information Engineering, Beijing Jiaotong University, Beijing 100044, China. (e-mail: renruixing0604@163.com; junhuizhao@hotmail.com)

He Fang is with the College of Computer and Cyber Security, Fujian Normal University, Fuzhou 200240, China (e-mail: fanghe@fjnu.edu.cn).

Xiaoke Sun is with the National Computer Network Emergency Response Technical Team/Coordination Center of China (CNCERT/CC), Beijing 100029, China (e-mail: xiaokesun@cert.org.cn).

Jiamin Li, Pengcheng Zhu, and Dongming Wang are with
the National Mobile Communications Research Laboratory, Southeast University, Nanjing 210096, China, and also with Purple Mountain Laboratories, Nanjing 211111, China (e-mail: jiaminli@seu.edu.cn; p.zhu@seu.edu.cn; wangdm@seu.edu.cn).
	
}

}

\markboth{IEEE Network, Special Issue, submission version}%
{Author \MakeLowercase{\textit{et al.}}: Defending Consensus-Based Cooperative Sensing in 6G UAV Swarms}

\maketitle

\begin{abstract}
In integrated sensing and communication (ISAC) enabled six-generation (6G) networks, unmanned aerial vehicle (UAV) swarms increasingly rely on consensus-based cooperative sensing: each node iteratively exchanges and fuses decisions with its neighbors to approach a global picture of the environment, supporting the high-reliability, low-latency inference demanded by mission-critical tasks. Yet this very mechanism of conformity becomes the lever of Byzantine attacks. By exploiting the update rule, malicious nodes manufacture a false consensus that lures honest nodes into drifting away from their true observations round after round, so that a normal node's actual error probability evolves dynamically and rises far above its intrinsic sensing error. Conventional fusion models, which assume node errors are fixed and independent, are therefore severely mismatched. This article argues for a different defensive perspective: instead of hunting for bad nodes, we seek to understand how good nodes turn bad, characterizing the inducement dynamics with evolutionary game theory and handing this law to the fusion center so that it can dynamically correct the true error level of normal nodes. Simulations show that the approach sustains near-perfect situational inference across a wide range of attack intensities and network sizes, while naturally supporting low-latency decisions. We further discuss several open challenges on the path toward secure ISAC-6G swarm sensing.
\end{abstract}


\IEEEpeerreviewmaketitle

\section{Introduction}
\IEEEPARstart{S}{ix} generation (6G) no longer treats communication and sensing as two separate matters. Under the integrated sensing and communication (ISAC) paradigm, a single wireless waveform both carries data and probes the environment \cite{ref1}, giving a unmanned aerial vehicle (UAV) swarm a compelling capability: to communicate with one another and cooperatively sense at the same time, distilling scattered observations into a collective judgment about the world \cite{ref2}. For tasks where a single mistake can be catastrophic, such as disaster search-and-rescue, low-altitude traffic supervision, border situational awareness, this high-reliability and low-latency cooperative sensing is exactly what the 6G vision (IMT-2030) promises \cite{ref3}.

The power of a swarm comes from coordination, and the underlying mechanism is surprisingly simple. Flocks of birds and schools of fish have long shown that if each individual simply watches its immediate neighbors and stays aligned with them, a vast group can remain orderly without any central command \cite{ref4}. Engineered UAV swarms inherit this wisdom, call it \emph{conformity}, aligning with neighbors and coordinating with the surroundings \cite{ref5}. More importantly, this idea has been embodied in concrete cooperative-sensing protocols: under consensus-based cooperative sensing and distributed detection, nodes not only sense on their own but also repeatedly exchange and weightedly fuse each other's decisions, progressively converging on a globally consistent conclusion, attaining near-centralized detection performance without any central fusion center \cite{ref6}. Conformity thus turns from a natural intuition into the engineering cornerstone of cooperative swarm sensing.

But it is precisely this cornerstone that hides a crack. Such a cooperative paradigm is, in essence, a consensus-based iterative decision process, and its openness makes it inherently vulnerable \cite{ref7}: a compromised UAV, a so-called Byzantine node, can not only falsify its own reports but, through the inter-node update rule, continuously inject erroneous values into its neighbors. Under ISAC, where sensing and communication share the same open spectrum, this attack surface is further enlarged. Conventional wisdom holds such attacks to be external and isolated: bad nodes each lie on their own, and the fusion center can pin them down with statistical anomaly detection or majority voting \cite{ref8}.

What this article argues is that, in a cooperative-sensing swarm built on neighbor trust and iterative conformity, matters are far from this simple. The very virtue of conformity can be turned against the swarm by an attacker. Building on this insight, the contributions of this paper are summarized as follows:
\begin{itemize}
	\item A novel class of threats, namely conformity-driven Byzantine attacks, is identified and elaborated. Malicious nodes exploit the update rules of consensus-based collaborative sensing to fabricate false consensus, thereby transforming the native collaborative mechanism of clusters into an error amplifier.
	\item An evolutionary game framework is adopted to characterize the evolutionary law governing the proportion of honest nodes diverted by adversarial inducement. This evolutionary law is then delivered to the fusion center, enabling dynamic calibration of the inherent drifting error rates of legitimate nodes.
	\item Simulation results verify that the proposed conformity-aware fusion scheme achieves near-perfect situational inference across a wide range of attack intensities and network scales, whereas conventional independent fusion fails completely under the same conditions.
\end{itemize}

The remainder of this paper is organized as follows. Section \ref{Sec2} elaborates on the dual-edged nature of conformity and the formation mechanism of false consensus. Section \ref{Sec3} illustrates the inherent mismatch of defense based on conventional independent fusion under such attacks. Section \ref{Sec4} presents the evolutionary game modeling of adversarial induction dynamics, the proposed conformity-aware robust fusion scheme derived therefrom. Section \ref{Sec5} conducts simulation-based validation of the proposed method. Section \ref{Sec6} discusses open challenges and future research directions. Section \ref{Sec7} concludes the entire paper.

\section{Conformity: A Double-Edged Sword} \label{Sec2}
Consider an honest UAV. Under a consensus-based cooperative-sensing protocol, each node generates its own sensing judgment, yet it adjusts its conclusion by referencing neighboring nodes’ decisions in accordance with the protocol’s update rules. In normal conditions this is beneficial, since fusing neighbor information is exactly how cooperative sensing improves accuracy. Now, what happens if a few malicious neighbors lurk nearby and consistently report the wrong information?

This honest UAV faces a dilemma: hold to its own judgment, or move toward the neighbors' majority? The update rule of cooperative sensing itself nudges it toward the latter. It begins to doubt itself and blend into the neighbors' erroneous conclusion, so that the decision it reports in the next round also departs from the truth. And its error, in turn, becomes the new majority that its own neighbors see in the following round. The lie thus spreads outward ring by ring, like ripples raised by a stone dropped into water.

\begin{figure}[!t]
\centering
\includegraphics[width=3.6in]{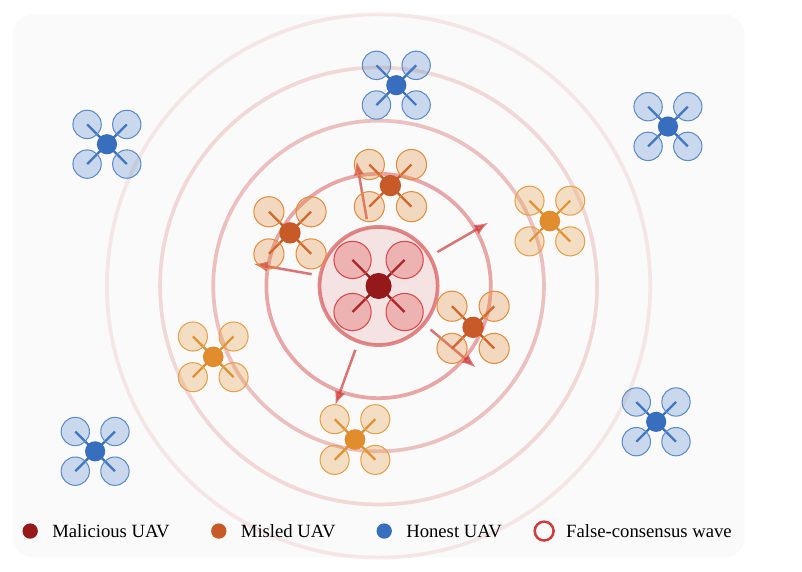}
\caption{Propagation of a false consensus. Malicious UAVs (red) create locally consistent erroneous reports; driven by the conformity rule, neighboring honest UAVs (blue) fall in line round after round, and the error ripples outward, steadily raising the actual error rate of normal nodes.}
\label{fig1}
\end{figure}

This is what we call a false consensus, as shown in Fig.~\ref{fig1}. Malicious nodes need not infect every UAV. They only have to create, locally, the illusion that everyone is saying so, and the conformity rule will carry out the rest of the propagation for them. A mechanism meant to help the swarm reach a correct consensus is hijacked to reach a wrong one. This is strikingly similar to how a rumor spreads layer by layer through conformity in a social network \cite{ref9}. The attacker uses not brute force, but the swarm's own logic.

This shift of perspective matters. In the past we regarded Byzantine nodes as noise sources, producing random, isolated errors \cite{ref7}. But in a conforming swarm, the real danger is not the handful of lies the bad nodes tell themselves, but the lies they induce the good nodes to tell collectively. The bad nodes are the spark; conformity is the dry tinder.

\section{Why Conventional Defenses Fail}\label{Sec3}
Mainstream Byzantine-resilient fusion methods are almost all built on two implicit assumptions: first, that a normal node's error probability is fixed (roughly equal to its sensor accuracy); and second, that errors across nodes are mutually independent. On these two premises, the fusion center can wield clean statistical models to screen out anomalies and vote \cite{ref8}. Prior work has also noted that Byzantine attacks in consensus-based sensing are not limited to data falsification but include consensus disruption---malicious nodes disobeying the update rule and arbitrarily manipulating their own values to disturb network-wide convergence \cite{ref10}. Even so, the defender typically still assumes normal nodes are honest and independent, with the attack coming only from those flagged bad nodes. Conformity breaks both assumptions at once.

Take fixed first. Under the inducement of a false consensus, a normal UAV's true error probability is no longer equal to its sensing accuracy. It varies dynamically with the density of malicious influence around it, and can climb far above its own sensing error. A fusion center that still models the error rate as constant and equal to sensor accuracy will grossly underestimate the true error level of normal nodes, and the model parts ways with the data. Existing resilient-consensus and defense methods largely still treat normal nodes as honest and independently erring \cite{ref11,ref12}, and thus never touch this dynamically evolving core.

Now take independent. Conformity means neighbors influence one another: what a UAV reports today depends on the neighbor reports it saw yesterday. Errors are no longer dice rolled in isolation, but a coupled, generation-to-generation chain reaction. To fuse them as independent events is to assume each at random about what is in fact an organized collusion and the result is a rout.

In a word, conventional methods thrive in a world where errors are fixed and independent, but a conformity-driven attack manufactures precisely a world where errors drift and couple. The tool is mismatched to the battlefield.

\section{Evolutionary-Game Modeling and Robust Fusion}\label{Sec4}
Since the root of the problem is the dynamic evolution of normal nodes after they are induced, the first step of defense should not be to rush to catch bad nodes, but to read the law of this evolution.

We borrow the tool of evolutionary game theory. It was originally devised to describe how a population of individuals gradually adjusts their strategies through repeated local interactions \cite{ref13}, which fits naturally with the watch the neighbors, adjust the self distributed decision-making of a UAV swarm. In our formulation, each normal UAV weighs between staying faithful to its own sensing and falling in with neighbors, while malicious neighbors keep exerting the pressure to fall in with me. Round after round, the fraction of the population that has been led astray converges toward a stable level---much as ripples eventually settle into a new water line. This convergence can be rigorously characterized by evolutionary dynamics on graphs \cite{ref14}.

\begin{figure}[!t]
	\centering
	\includegraphics[width=3.4in]{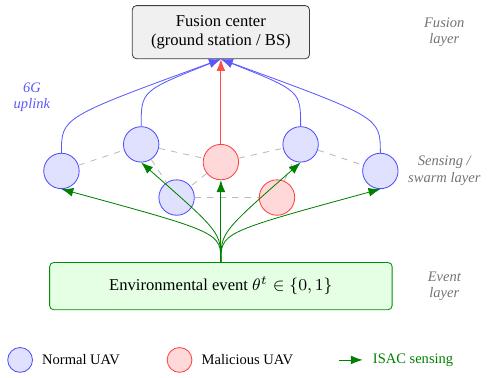}
	\caption{System architecture of ISAC-6G UAV swarm cooperative sensing. UAVs sense the environmental event with ISAC signals and report local decisions to the fusion center over 6G uplinks; the conformity interaction among neighboring UAVs is exploited by malicious nodes to propagate inducement.}
	\label{fig2}
\end{figure}

Crucially, this law is not merely a final resting point but a predictable trajectory over time. From an initial state in which no honest node has yet been misled, the misled fraction can be propagated forward step by step: at every time instant, the fusion center can compute in advance roughly how large a share of the normal nodes has already drifted. What used to be an unknown, silently growing corruption thus becomes a quantity the defender can track in real time. This trajectory, moreover, has a clear dependence on three factors: sensing channel quality (the clearer a UAV sees, the harder it is to lead astray), the fraction of malicious nodes (the more bad neighbors, the more severely it is misled), and attack intensity (the more extreme the bad nodes, the fiercer the inducement). In other words, we no longer treat how much a normal node errs as an unknown black box, but turn it into a computable, predictable quantity.

With this yardstick in hand, the second step of defense follows naturally: hand the evolution law to the fusion center. Concretely, the fusion center processes the incoming reports in three stages. First, instead of assuming that a normal node errs at its outdated sensor accuracy, it uses the predicted misled fraction at the current instant to estimate the true, drifting error level of normal nodes. Second, guided by this estimate, it assigns each report its proper credibility: at instants when honest nodes are more likely to have been led astray, the weight placed on a normal-looking report is adjusted accordingly, so that induced errors are no longer mistaken for trustworthy evidence. Third, combining this with the known statistical signature of malicious nodes, it renders a maximum-a-posteriori decision on the true situation. The whole procedure needs only the model parameters and the reports themselves, no ground truth about who is malicious is ever required.

Counterintuitively, the harder the attack, the better this works. A fiercer inducement drives a more pronounced, and therefore more predictable, evolution trajectory; the fusion center can then resolve which errors were induced and subtract them out all the more cleanly. This is precisely why, as the simulations will show, the defense grows more rather than less reliable as the attack intensifies.

The key shift here is this: we did not build a more elaborate bad-node detector, but gave the fusion center a model of how the population drifts under inducement. What it defends against is not a few bad nodes, but the entire induced, collective drift. The composition of the whole system UAVs cooperatively sensing with ISAC signals, reporting over 6G uplinks, and the fusion center performing robust inference guided by the evolution law is shown in Fig.~\ref{fig2}.

\section{Simulation Verification}\label{Sec5}
\begin{figure}[!t]
\centering
\includegraphics[width=3.4in]{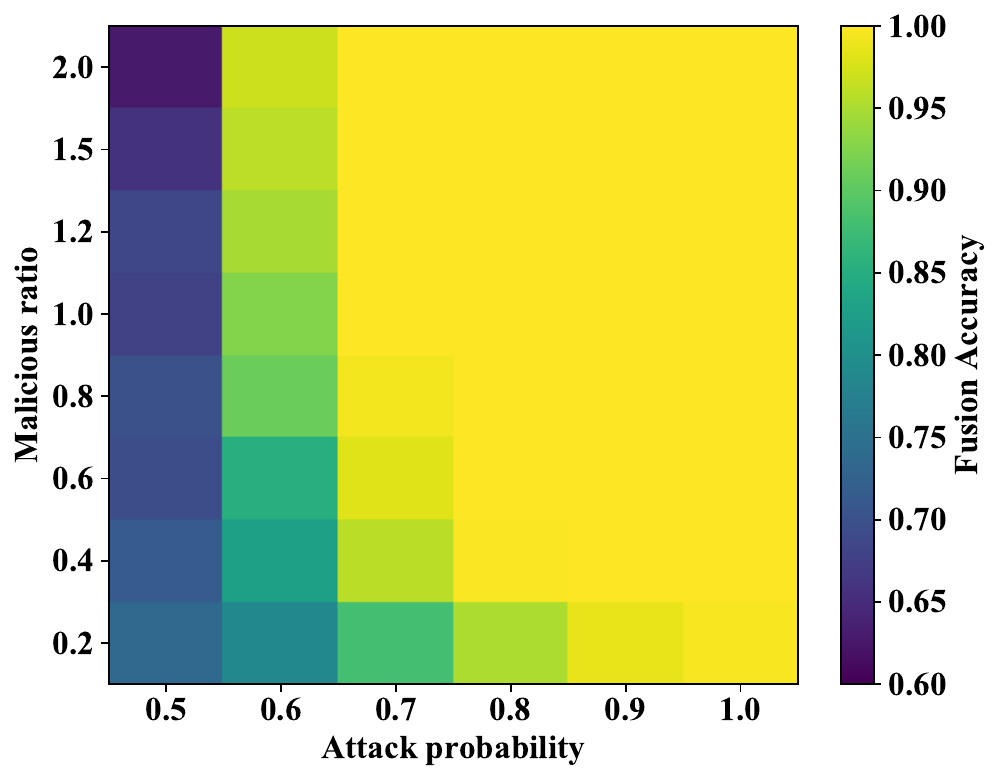}
\caption{Attack-intensity phase map. Over the plane spanned by malicious ratio and attack probability, the fusion accuracy of the proposed mechanism is near-perfect across most of the region, dropping only in the most extreme corner.}
\label{fig3}
\end{figure}

\begin{figure}[!t]
\centering
\includegraphics[width=3.4in]{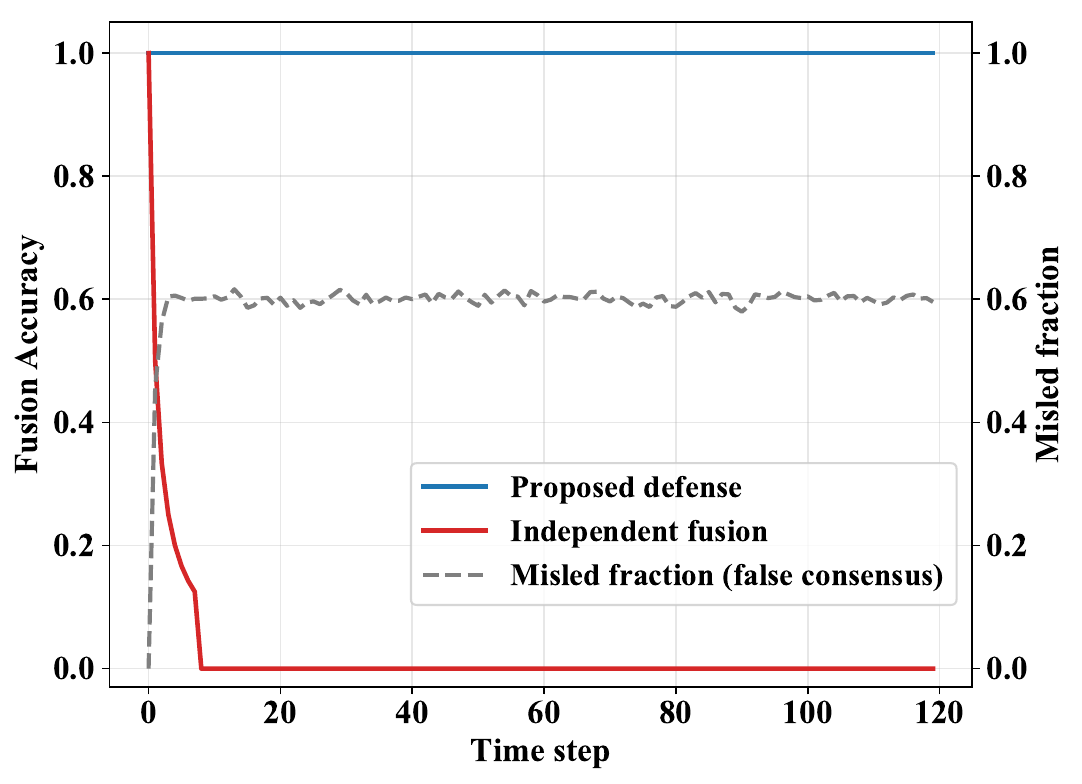}
\caption{False-consensus propagation vs.\ defense interception. Over time, the fraction of misled normal nodes rises quickly to about sixty percent (false-consensus spread); conventional independent fusion collapses accordingly, while the proposed mechanism sustains near-perfect situational inference.}
\label{fig4}
\end{figure}

We validate this idea through extensive Monte Carlo simulations, and select here the two groups of results that speak most clearly. Unless otherwise stated, the simulations use a regular network of three to five hundred UAVs, each connected to ten neighbors, and every data point is averaged over multiple Monte Carlo runs. To confirm that the findings are not tied to a particular topology, scale-free and random networks are also tested, yielding consistent trends. The intrinsic sensing error is kept at a moderate level, and the fusion window is short so as to reflect low-latency operation. In the phase map of Fig.~\ref{fig3}, the malicious ratio and the attack probability are swept over their respective ranges; in the propagation experiment of Fig.~\ref{fig4}, a strong attack is applied so that the spread of the false consensus is clearly visible.

A Byzantine attack has two knobs: the number of malicious nodes (ratio) and how aggressively they lie (attack probability). We spread these two knobs into a plane and check, cell by cell, the situational-inference accuracy of the proposed mechanism. As shown in Fig.~\ref{fig3}, the result is a broad oasis of high accuracy across the vast majority of attack combinations, fusion accuracy stays close to 100\%; only in the most extreme corner, where malicious nodes reach twice the number of normal nodes and the attack becomes semi-random, does it slacken. This shows the defense is effective not merely at some cherry-picked operating point, but is broadly robust over a wide range of attack intensities, exactly what a 6G network facing real, shifting threats requires.

The lie spreads, yet the defense holds. This group of results echoes the ripple narrative of Fig.~\ref{fig1} directly. As shown in Fig.~\ref{fig4}, we let a strong attack ferment within the swarm and track two things unfolding at once. First, the fraction of misled normal nodes climbs over time, rising rapidly from zero to approximately sixty percent. This faithfully depicts how false consensus propagates layer by layer and drags the majority of honest nodes into erroneous judgments. Second, how the two fusion mechanisms perform amid this spread. Conventional independent fusion, which assumes fixed and independent errors, collapses as the false consensus spreads, its accuracy plunging toward zero---it is utterly drowned by the lie. The proposed mechanism, by contrast, holds near-perfect accuracy throughout: even though more than half the nodes have been induced, its grasp of the evolution law lets it keep correcting the contaminated information back to the truth. The same attack, one drowned and one unmoved. This vividly confirms the value of the understand the evolution idea.

It is also worth noting that this fusion mechanism naturally supports low-latency decisions. It renders independent judgments at each time instant and conducts online prediction of the instantaneous error level, thus requiring no accumulation over an extended observation horizon. Even the shortest fusion window can deliver near-perfect accuracy and minimize decision latency. This is especially important for mission-critical 6G applications.

\section{Open Challenges and Outlook}\label{Sec6}
Taking understanding population evolution as the handle for swarm security is only a starting point. Several directions merit deeper cultivation on the way to truly practical ISAC-6G swarm security.

\emph{From the sensing layer to the waveform layer.} This article characterizes attack and defense at the level of sensing decisions. One layer deeper, the ISAC attack surface actually begins at the shared waveform itself, since malicious nodes may contaminate echoes or forge sensing signals at the physical layer. \cite{ref7}. Pushing the evolutionary perspective down to the waveform and signal-processing layer is a valuable open problem.

\emph{Bringing latency and reliability into the loop.} 6G promises guaranteed low latency and ultra-high reliability. How to explicitly incorporate the decision latency of the fusion mechanism, and its reliability under attack, as quantifiable, optimizable network metrics within the design loop, deserves systematic study \cite{ref15}.

\emph{From static laws to adaptive adversaries.} Real-world attackers change tactics. When malicious nodes dynamically adjust their inducement strategy, the defender's evolution model must also adapt online, pushing the problem toward a dynamic game between attacker and defender, a far broader battlefield.

\emph{Toward standardization and heterogeneous cooperation.} UAV swarms will ultimately connect into larger heterogeneous 6G networks (terrestrial base stations, non-terrestrial platforms, multiple sensing modalities). How conformity-driven attacks behave in heterogeneous, cross-domain settings, and the standardization of the corresponding defenses, is a necessary path to real deployment.

\section{Conclusion}\label{Sec7}
This article studied the security of consensus-based cooperative sensing in ISAC-enabled 6G UAV swarms. A conformity-driven Byzantine attack was identified, in which malicious nodes exploit the neighbor update rule to manufacture a false consensus, so that the induced errors of honest nodes evolve dynamically rather than staying fixed and independent, rendering conventional independent fusion severely mismatched. To counter this, the inducement process was modeled via evolutionary game theory, and the resulting law was embedded into a conformity-aware fusion mechanism that dynamically corrects the true error level of normal nodes. Simulations confirmed near-perfect situational inference across diverse attack intensities and network sizes, clearly outperforming independent fusion at low decision latency. These results suggest that understanding the collective evolution of induced nodes, rather than merely detecting individual attackers, is an effective route to secure cooperative sensing in open-airspace 6G swarms.

\end{document}